\documentclass[preprint,12pt]{elsarticle}

\usepackage{amssymb}

\usepackage{anysize}
\marginsize{1cm}{1cm}{1cm}{1cm}

\begin{document}

\begin{frontmatter}

\title{Equilibrium vortex lattices of a binary rotating atomic Bose-Einstein condensate with unequal atomic masses}


\author{ Biao Dong  \fnref{label1,label2}}
\author{ Lin-Xue Wang \fnref{label1,label3}}
\author{ Guang-Ping Chen \fnref{label1,label2}}
\author{ Wei Han \fnref{label1}}
\author{ Shou-Gang Zhang \fnref{label1}}
\author{ Xiao-Fei Zhang \fnref{label1}}

\address[label1]{Key Laboratory of Time and Frequency Primary Standards, National Time Service Center, Chinese Academy of Sciences, Xi'an 710600, China}
\address[label2]{University of Chinese Academy of Sciences, Beijing 100049, China}
\address[label3]{College of Physics and Electronic Engineering, Northwest Normal University, Lanzhou 730070, China}

\begin{abstract}

We perform a detailed numerical study of the equilibrium ground-state structures of a binary rotating
Bose-Einstein condensate with unequal atomic masses. Our results show that the ground-state distribution
and its related vortex configurations are complex events that differ markedly depending strongly
on the strength of rotation frequency, as well as on the ratio of atomic masses. We also discuss the
structure and radius of the clouds, the number and the size of the core region of the vortices, as a
function of the rotation frequency, and of the ratio of atomic masses, and the analytical results
agree well with our numerical simulations. This work may open an alternate way in the quantum control
of the binary rotating quantum gases with unequal atomic masses.

\end{abstract}

\end{frontmatter}

\section{Introduction}

Vortices are fundamental excitations of nonlinear media and have attracted great interest in
diverse contexts of science and engineering, such as superconductor, nonlinear optics, and
recently in Bose-Einstein condensate (BEC) \cite{G. V. Chester,R. E. Packard,E. Lundh,R. Dum,M. R. Matthews,
J. E. Williams,S. Stringari,J. R. Abo-Shaeer,A. E. Leanhardt,X. F. Zhang,J. Lovegrove}.
Due to the existence of macroscopic wave function of the condensate, the realization of BEC has brought
about a suitable research environment for investigating the general properties of superfluidity
and superconductivity with a high degree of control and versatility, which manifest themselves
as the presence of quantized vortices \cite{K. W. Madison,M. Ueda}. When the rotation frequency exceeds a critical one,
vortices begin enter the system and form a vortex lattice. It has also been well established
that the ground state of such a system under a rapid rotation is a vortex lattice, and such lattices
containing hundreds of vortices have been observed in experiments \cite{A. L. Fetter,P. Kuopanportti}.

With the recent prospects of producing multi-component condensate, in which the mixture has been
realized by using distinct elements \cite{G. Ferrari,G. Modugno,G. Thalhammer,J. Catani2,A. Lercher,D. J. McCarron,B. Pasquiou}, different isotopes of the same element \cite{S. B. Papp,S. Sugawa}, or different internal spin states of the same
isotope \cite{C. J. Myatt,D. S. Hall,G. Delannoy,V. Schweikhard,K. Kasamatsu,R. P. Anderson},
vortex states in multi-component condensates have become even more important. The lattice
structures of such a system can---contrary to its single-component counterpart where the lattice
structure is always a triangular lattice---contain a myriad of unusual ground-state vortex
structures, and can be obtained either by increasing the number of components or by introducing
long-range interactions \cite{K. Kasamatsu}. More specially, for a mixture superfluid consisting
of two-component BEC, which is described by two $\mathbb{C}$-valued order parameters $\Psi_1$ and
$\Psi_2$ \cite{X. F. Zhang2}, a rich variety of exotic vortex structures,
such as coreless vortices, giant skyrmions, interlacing square vortex lattices, and serpentine
vortex sheets, have been predicted \cite{E. J. Mueller,K. Kasamatsu1,S.-J. Yang,K. Kasamatsu2}.

In a recent theoretical work, Kuopanportti {\it et al}. have investigated the exotic vortex lattices
of such a binary system with unequal atomic masses and attractive intercomponent interaction, where lattice having a
square geometry or consisting of two-quantum vortices have been found \cite{P. Kuopanportti}. However,
the ground-state and rotational properties of such a binary BEC with unequal atomic masses has not been
studied as thoroughly \cite{B. D. Esry,R. Ejnisman,H. Pu,X. Liu,P. F. Bedaque,A. Alonso-Izquierdo,P. Kuopanportti2}.
To our knowledge, there has been little work on the size of the condensate, and on the number and the size
of the core region of the vortices, which is what we attempt to do in this work. In this paper, we carry out
a detailed numerical analysis of the combined effects of the ratio of atomic masses and rotation frequency
on the ground-state vortex structures of a binary rotating atomic BEC with unequal atomic masses.
Our results show that the ground-state distribution and its related vortex configurations are complex events that
differ markedly depending strongly on the strength of rotation frequency, as well as on the ratio of atomic masses.

The rest of this work is organized as follows. In Sec. 2 we formulate the theoretical
model describing a binary BEC with unequal atomic masses, and briefly introduce
the numerical method. The ground-state density distributions and its related vortex structures,
together with the number and the size of the core region of the vortices, as a function
of the ratio of atomic masses, and of the rotation frequency, are discussed in Sec. 3.
Finally, in Sec. 4, we summarize the main results of the work.

\section{Theoretical model and method}

Generally speaking, for different atomic species, the critical temperatures below which
BEC take place are different. Thus we consider the lower critical
temperature $T_c= \min \{ T_{c1}, T_{c2} \}$, at which both phase transitions take place.
Within the framework of zero-temperature mean-field theory, where the collisions between
the condensate atoms and the thermal cloud are neglected, the ground-state and dynamics
of a binary (or two-component) BEC are described in terms of two complex-value order-parameters
$\Psi_{1}$ and $\Psi_{2}$. For a quasi-two-dimensional (Q2D) system and viewed from the frame
of reference co-rotating with the trapping potential, the Gross-Pitaevskii (GP) energy function of such a system
can be written as \cite{X. F. Zhang2},

\begin{eqnarray}
E[\Psi_1,\Psi_2]=  \int dr \biggl \{\sum_{i=1,2} \Psi_i^{*}\big (-\frac{\hbar^2\nabla^2}{2m_i}+
V_{i}(\textbf{r})+\frac{U_{ii}}{2}|\Psi_i|^2 -\Omega_i L_z \big )\Psi_i+U_{12}|\Psi_1|^2|\Psi_2|^2 \biggl \},
\end{eqnarray}
where $m_i$ is the mass of the boson $i$, and the rotating term is given by $\Omega_i L_{z}=i\hbar
\Omega_i (y\partial x-x\partial y)$. The external trapping potential is assumed to be $V_{i}
(\textbf{r})=m_i[\omega_{ix}^2 x^2 +\omega_{iy}^2 y^2 + \omega_{iz}^2 z^2]/2$, with $\omega_{ix}$,
$\omega_{iy}$, and $\omega_{iz}$ being the $x$-, $y$-, and $z$- directions trapping frequencies of the
harmonic potential. The coefficients $U_{ii}=4\pi\hbar^2a_{ii}/m_i$ and $U_{12}=U_{21}=2\pi\hbar^2a_{12}/m_R$
denote the intra- and inter-component interactions, which are related to the $s$-wave scattering lengths
$a_{ii}$ and $a_{12}$ respectively, with the reduced mass $m_R =  m_1m_2/(m_1+m_2)$. The Q2D condensate
obtained by adding a very strong confinement along $z$ axis, {\it i.e.}, $\hbar \omega_{iz}$ being much
larger than any other energy scale in the problem, allows us to separate the order parameter in the form
$\Psi_i(r)=Z_i(z) \psi_i(x,y)$, with $Z_i(z)=e^{-z^2/2a_{zi}^2}/(\pi a_{zi}^2)^{1/4}$ denoting
the ground state of the harmonic oscillator with frequency $\omega_{iz}$ and $a_{zi}=[\hbar/(m_i\omega_{iz})]^{1/2}$.

By using the variational procedure \cite{V. M. Perez-Garcia,W. Bao2}, $i\hbar \partial \psi_i = \delta E / \delta \psi^{*}_i$,
we arrive at a pair of dimensionless \cite{M. J. Holland,F. Dalfovo,F. Dalfovo2} coupled GP equations,

\begin{eqnarray}
i\frac{\partial\psi_1}{\partial t}  &= &  \biggl( -\frac{\nabla^2}{2\gamma}+ \gamma V_1 + \!\sqrt{\frac{m_2}{\gamma}}\textmd{g}_{11}|\psi_1|^2 +\sigma\textmd{g}_{12}|\psi_2|^2 - \Omega L_z \biggl )\psi_1 ,\nonumber\\
i\frac{\partial\psi_2}{\partial t} &= & \biggl (-\frac{\nabla^2}{2}+ V_2 +  \sigma\textmd{g}_{21}|\psi_1|^2 +\!\sqrt{m_2}\textmd{g}_{22}|\psi_2|^2 - \Omega L_z \biggl )\psi_2, \label{CGP}
\end{eqnarray}
where the mass ratio is defined as $\gamma=m_1/m_2$ and $\sigma=m_2/\sqrt{2m_R}$ for the facilitation of expressions. And $\textmd{g}_{ii} \!= \!
4\pi a_{ii}\sqrt{1/(2\pi\hbar)}$ and $\textmd{g}_{12} \!=\textmd{g}_{21}= \! 4\pi a_{12}\sqrt{1/(2\pi\hbar)}$ are the
intra- and inter-species coupling constants. Here we specialize to the ``balanced" case with
$N_1=N_2=N$, $\omega_{m}=\min \{\omega_{ix},\omega_{iy} \}=\omega_{ix}=\omega_{iy}=1$, $\Omega_1=\Omega_2=\Omega$ and $\textmd{g}_{11}=\textmd{g}_{22}=\textmd{g}$, and
work in dimensionless units by scaling with the trap energy $\hbar \omega_m$ and the characteristic oscillator length $a_0=\sqrt{ \hbar / m_2\omega_m}$,
where appropriate. Thus, the external trapping potential $V_1=V_2=(x^2+y^2)/2$, and the order parameters are normalized as $\int|\psi_1|^2d\textbf{r}=\int|\psi_2|^2d\textbf{r}=1$.

To obtain the ground-state wave functions of such a system, one can minimize the total energy function of
such a system with respect to $\psi_i$, keeping the total number of particles fixed. Thus, the stationary
modes (the so-called time-independent version) of Eq. \ref{CGP}, with chemical potential $\mu_i$ corresponding
to $\psi_i (\textbf{r},t)= \phi_i (\textbf{r})\exp(-i \mu_i t/\hbar)$ \cite{L. P. Pitaevskii}, can be written as

\begin{eqnarray}
\mu_1\phi_1 &= & \biggl( -\frac{\nabla^2}{2\gamma}+ \gamma V_1 + \!\sqrt{\frac{m_2}{\gamma}}\textmd{g}_{11}|\phi_1|^2 +\sigma\textmd{g}_{12}|\phi_2|^2
- \Omega L_z \biggl )\phi_1 ,\nonumber\\
\mu_2 \phi_2 &= & \biggl (-\frac{\nabla^2}{2}+ V_2 +  \sigma\textmd{g}_{21}|\phi_2|^2 +\!\sqrt{m_2}\textmd{g}_{22}|\phi_1|^2
-\Omega L_z \biggl )\phi_2. \label{SCGP}
\end{eqnarray}

To further simplify the numerical simulations, and highlight the effects of the ratio of
atomic masses and the rotation frequency, we further fix the intra-species interaction
$\textmd{g}=100$ and inter-species interaction $\textmd{g}_{12}=20$. The equilibrium vortex structures of the system in different parameter
space are obtained by using the Backward Euler pseudospectral (BESP) scheme within an imaginary-time
propagation approach \cite{W. Bao,X. F. Zhang3,T. Mithun,X. Antoine2,X. Antoine}. For the initial
data, we prepare the initial wave function as a combination of the ground state of harmonic potential
and a vortex state for the non-interacting case, which can be written as

\begin{equation}
\phi(x,y,0)=\frac{(1-\Omega)\phi_{h0}(x,y)+\Omega\phi_{h0}^\upsilon(x,y)}{\||(1-\Omega)\phi_{h0}(x,y)+\Omega\phi_{h0}^\upsilon(x,y)\||_0},
\end{equation}
with
\begin{eqnarray}
&\phi_{h0}&(x,y)=\frac{1}{\sqrt{\pi}}e^{-(\gamma_xx^2+\gamma_yy^2)/2},\nonumber\\
&\phi_{h0}^\upsilon&(x,y)=\frac{\gamma_xx+i\gamma_yy}{\sqrt{\pi}}e^{-(\gamma_xx^2+\gamma_yy^2)/2},
\end{eqnarray}
where $\gamma_x=\omega_x/\omega_m$ and $\gamma_y=\omega_y/\omega_m$. We stress that the
obtained vortex structures should depend sensitively on the initial
wave functions in different parameter regions. In the following numerical simulations, our numerical results
show that such initial data always gives the real ground state as long as the rotation frequency is not
too large, {\it i.e.,} $\Omega  <  \omega_m$ for the Q2D isotropic harmonic potential considered here.

\section{Results and discussion}

\subsection{Numerical Results}

As is referred before, the rotation frequency and the ratio of atomic masses play a very important role
on the ground-state and rotational properties of the system. In what follows, we will perform a series of
numerical experiments to study the combined effects of rotation frequency and the ratio of atomic masses
on the structure and radius of the clouds, and on the number and the size of the core region of the vortices.

Fig. \ref{F1} exhibits the typical density and phase distributions of the system for fixed
rotation frequency $\Omega=0.6$, but for different values of the ratio of atomic masses $\gamma=1,2,4,9$,
corresponding to (a), (b), (c), and (d), respectively. In this case, due to the small rotation and
our choice of the initial data, a single central vortex is generated for both components $1$ and $2$.
It is clear from Fig. \ref{F1} that such density distributions depend sensitively
on the ratio of atomic masses. Upon increasing $\gamma$, the ground-state density distributions
of component $1$ show overall concentrated contraction tendency, with the central vortex structure
unchanged. However, the sizes of the core region of the central vortices contract as the ratio
of atomic masses increasing. For component $2$, the sizes of the core region, the density
distributions and the related vortex structures are essentially not affected by such changes.

Shown in Figs. {\ref{F2}-\ref{F4}} are the density and phase distributions of the system for
the same values of the ratio of atomic masses as in Fig. \ref{F1}, but for different rotation
frequencies $\Omega=0.7,0.8,0.9$, corresponding to Figs. {\ref{F2}-\ref{F4}}, respectively.
For a relatively larger rotation frequency $\Omega=0.7$, it is observed that a triangular vortex
lattice is formed for each component in the case of equal masses ($\gamma=1$, see Fig. \ref{F2}(a)).
As the ratio of atomic masses increasing, the ground-state density distribution of component $1$
shows overall concentrated contraction tendency, accompanied with a decrease of the sizes of the core region of the vortices. Typical density and phase distributions of such case are shown in Figs. \ref{F2} (b)-(d). We note that the overall
concentrated contraction tendency of component $1$ also appears in the following discussions and
in what follows we will give a more detailed physical explanation.

When the rotation frequency is further increased to $\Omega=0.8$ and $ 0.9$ (as shown in
Figs. \ref{F3} and \ref{F4}, respectively), the number of vortices of component $2$ basically increases
with the ratio of atomic masses, on one hand; and on the other hand, the new nucleated vortices,
together with the previous ones, form a vortex necklace structure at the edge of the cloud.
More interestingly, in the central region of the cloud, we observe the formation of a single
central vortex (see Figs. \ref{F3} (b) and (d)) and triangular vortex lattice structures (see Figs. \ref{F4} (c) and (d)) for different parameter sets. Actually, this symmetric and central vortex structure
was discussed previously for a single-component BEC \cite{W. Bao}; and the explanation of such
structure lies the fact that it is energetically favorable for the vortices to site in the low-density
region and this special type of vortex structure is more beneficial to lower the total energy of the
system. While for component $1$, the number of vortices basically keep unchanging with the ratio of
atomic masses $\gamma$, accompanied with a structure change of the cloud. Although it is not yet
certain what mechanism is responsible, we may gain some physical insight by understanding the mass
difference between such two components, as discussed in the following subsection for the analytical results.

\subsection{Analytical Results}

Generally speaking, the ground-state density distributions of component $1$ show overall concentrated
contraction tendency upon increasing $\gamma$ in Figs. {\ref{F1}-\ref{F4}}. This phenomenon can be
easily explained by the change of chemical potential $\mu$ and the harmonic oscillator length $a_\perp=[\hbar/(m\omega_\perp)]^{1/2}$
in the transverse $x$-$y$ plane. The former one can be described by atomic mass $m$ and an effective interaction $\textrm{g}$
(which is fixed in our scheme) within the mean-field approach at low energy under Thomas-Fermi (TF) approximation
\cite{E. Lundh,F. Dalfovo2}. Thus, we can derive the relation between the radius of the condensate and the variable atomic
mass in the limit of large $N$ and TF approximation.


For the case of atomic species with a repulsive interaction, high particle numbers and not too strong
confinement, {\it e.g.}, $N|a_s|/a_{ho} \gg 1$, the condensate wave function is essentially dominated by
the interaction energy. By using the stationary version of GP equation (\ref{SCGP}) and the TF approximation
 neglecting the kinetic energy (the so-called quantum-pressure term
$(\textbf{p}-m_i {\bf{\Omega}} \times \textbf{r})^2 /2m_i)$ \cite{U. R. Fischer,C. Cohen-Tannoudji}, we obtain

\begin{eqnarray}
\mu_1=V_1-\frac{m_1}{2}({\bf{\Omega}} \times \textbf{r})^2 +\frac{U_{11}}{\sqrt{2\pi}a_{z1}}|\psi_1|^2 +\frac{U_{12}}{\sqrt{\pi(a^2_{z1}+a^2_{z2})}}|\psi_2|^2,\nonumber\\
\mu_2=V_2-\frac{m_2}{2}({\bf{\Omega}} \times \textbf{r})^2 +\frac{U_{22}}{\sqrt{2\pi}a_{z1}}|\psi_2|^2 +\frac{U_{21}}{\sqrt{\pi(a^2_{z1}+a^2_{z2})}}|\psi_1|^2.
\end{eqnarray}
The boundaries of the clouds are therefore given by

\begin{equation}
\mu_i=V_i(\textbf{r})-\frac{m_i}{2}({\bf{\Omega}} \times \textbf{r})^2.
\end{equation}
Thus one obtains the transverse radii of the clouds

\begin{equation}
R_i=\sqrt{\frac{2\mu_i}{m_i(\omega_i^2-\Omega^2)}}.
\end{equation}
By using the normalization constraint $\int|\psi_1|^2d\textbf{r}=\int|\psi_2|^2d\textbf{r}=N$, one derives

\begin{equation}
\mu_i^2=\frac{1}{\pi}\biggl[\frac{U_{ii}}{\sqrt{2\pi}a_{zi}} +\frac{U_{ij}}{\sqrt{\pi(a^2_{z1}+a^2_{z2})}}\biggl]N m_i(\omega_i^2-\Omega^2).
\end{equation}

In the limit of large $N$, the transverse radii of the clouds is given by \cite{C. J. Pethick,G. Baym}

\begin{equation}
\frac{R_i}{a_{i\perp}}= \biggl( \frac{2\mu_i\omega_\perp}{\hbar\omega_{ho}^2}\biggl)^{\frac{1}{2}},
\end{equation}
where the geometric average of the oscillator frequencies $\omega_{ho}=(\omega_x\omega_y\omega_z)^{1/3}$.
As a consequence, the relation between radii and masses leads to the overall
concentrated contraction tendency clear. For a binary BEC, we can approximately obtain the radius ratio
of such two species,

\begin{equation}
\frac{R_1}{R_2}=\biggl[\frac{10\sqrt{m_1m^4_2}+\sqrt{2(m^2_1m^3_2+m_1m^4_2)}}
{10\sqrt{m^4_1m_2}+\sqrt{2(m^3_1m^2_2+m^4_1m_2)}}\biggl]^{\frac{1}{2}}, \label{R}
\end{equation}


To give a qualitative explanation of the relation between the radius and the atomic mass of the condensate, the classical centripetal force can be employed. As shown in Fig. \ref{F5}, when one rotates a binary system (containing two species with unequal atomic masses) with fixed torque $\bf{\tau}=\textbf{r}\times \textbf{F}$, where $\textbf{F}=m\bf{\Omega}\times(\bf{\Omega}\times \textbf{r})$ represents the centripetal force, the bosons move as two clouds with different radii. Since $\bf{\tau}$ is fixed,
the radius of the cloud is smaller for one with bigger atomic mass. Actually, the centripetal force employed here is, to some extent, similar to the Magnus force, as discussed in \cite{R. Barnett1,R. Barnett2}.

In addition, according to the Feynman relation, it is easy to deduce the number of the vortices in each
component, which can be written as \cite{L. J. Campbell,E. J. Yarmchuk},

\begin{equation}
N_v(\Omega)\approx m\Omega R^2,
\end{equation}
then the size of the core region for the vortices is given by \cite{U. R. Fischer,I. Coddington},

\begin{equation}
r_i=\sqrt{\frac{\hbar}{m_i\Omega}}. \label{SI}
\end{equation}

To give a clear comparison with the numerical results, the radius ratio and the size ratio of a binary
rotating atomic BEC as a function of the ratio of atomic masses are shown in Figs. \ref{F6} and \ref{F7},
respectively, where blue rings correspond to the numerical results read from Figs. {\ref{F1}-\ref{F4}} for $\gamma=1,2,4,9$,
respectively. From these pictures, we can clearly see that our numerical results match well with the analytical
formula (\ref{R}) and (\ref{SI}).


Given the above analysis, we conclude that the ratio of atomic masses, together with the
rotation frequency, can be used not only to manipulate the structure and radius of the clouds,
but also to control the number and the size of the core region of the vortices.

Before the conclusion, let us consider the possibility of experimental realization of such
ground-state density distributions and its related vortex structures. In real experiments,
$\gamma=m_1/m_2$ is rarely strictly integer-valued but often sufficiently close to an integer
\cite{P. Kuopanportti}. Perhaps the most promising system for observing the described ground-state
structures in experiments is a binary atomic BEC, which consists of $^{41}$K$-$$^{87}$Rb ($^{23}$Li$-$$^{87}$Rb)
atoms, for the case of $\gamma=2$ ($\gamma=4$).

\section{Conclusion}

In summary, we have investigated the equilibrium vortex lattices of a binary rotating atomic
BEC with unequal atomic masses. Within the framework of mean-field theory, the ground-state
vortex structures are obtained by using the BESP scheme within an imaginary-time propagation
approach. Moreover, the combined effects of the ratio of atomic masses and rotation frequency on
the ground-state and rotational properties are investigated.

Our results show that the ground-state density distributions and its related vortex configurations
have a strong dependence on such system parameters. Upon increasing the ratio of atomic masses
$\gamma$ and the rotation frequency, the ground-state density distributions of component $1$
(massive atom) show overall concentrated contraction tendency, while the number of vortices
of such component keep unchanging; for component $2$, the number of vortices increases with the
ratio of atomic masses, on one hand; and on the other hand, the new nucleated vortices, together
with the previous ones, form a vortex necklace structure at the edge of the cloud. In addition,
in the central region of component $2$, we observe the formation of a single central vortex and
triangular vortex lattice structures for different parameter sets. Finally, we also discuss the
structure and radius of the clouds, the number and the size of the core region of the vortices,
as a function of the rotation frequency, and of the ratio of atomic masses, and the analytical
results agree well with our numerical simulations. The results not only help us better understand
the effects of the ratio of atomic masses on the ground-state structures, but also offer us an alternative
way to manipulate such two-component system.

{\bf Acknowledgements} We would like to express our sincere thanks to X. Antoine for valuable comments in
the treatment of the numerical code, and to Tao Zhang for useful discussions. This work was
supported by NSF of China.
X. F. Zhang is also supported by the key project fund of the CAS ``Light of West China" Program.

\newpage
\textbf{Figure Captions}

Fig. 1. (Color online) The ground-state density and phase distributions of a
binary rotating atomic BEC in an isotropic harmonic potential, for rotation frequency $\Omega=0.6$,
$\textmd{g}=100$, $\textmd{g}_{12}=20$, and for $\gamma=1,2,4,9$, corresponding to (a), (b), (c),
and (d), respectively. The third and fourth columns are the corresponding phase distributions.

\begin{figure}[tbp]
\centering
\includegraphics[height=6.5cm,width=6.5cm]{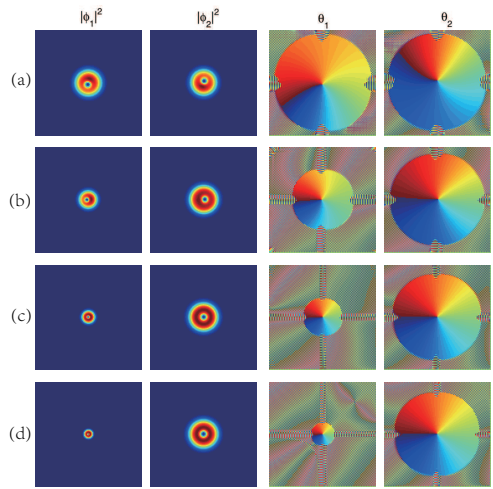}
\caption{(Color online) The ground-state density and phase distributions of a
binary rotating atomic BEC in an isotropic harmonic potential, for rotation frequency $\Omega=0.6$,
$\textmd{g}=100$, $\textmd{g}_{12}=20$, and for $\gamma=1,2,4,9$, corresponding to (a), (b), (c),
and (d), respectively. The third and fourth columns are the corresponding phase distributions.} \label{F1}
\end{figure}

Fig. 2. (Color online) The same as in Fig. {\ref{F1}} but for $\Omega=0.7$.

Fig. 3. (Color online) The same as in Fig. {\ref{F1}} but for $\Omega=0.8$.

Fig. 4. (Color online) The same as in Fig. {\ref{F1}} but for $\Omega=0.9$.

Fig. 5. (Color online) The qualitative explanation of the relation between the radius and the atomic mass of the condensate.
When one rotate a system mixing two species with unequal atomic masses with fixed
torque $\bf{\tau}=\textbf{r}\times \textbf{F}$,
where $\textbf{F}=m\bf{\Omega}\times(\bf{\Omega}\times \textbf{r})$ represents the centripetal force,
the bosons move as two clouds with different radius.

\begin{figure}[tbp]
\centering
\includegraphics[height=6.5cm,width=6.5cm]{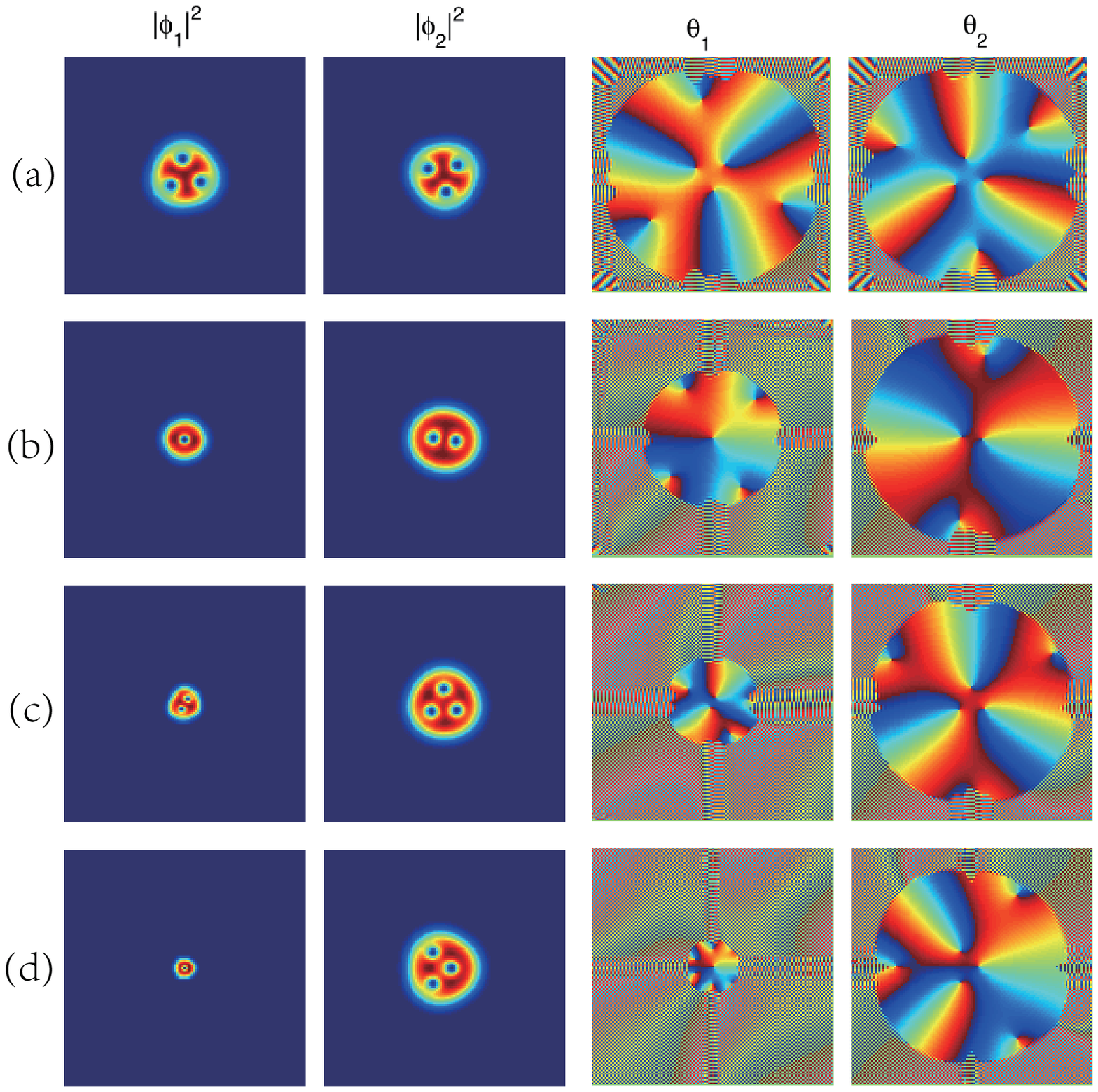}
\caption{(Color online) The same as in Fig. {\ref{F1}} but for $\Omega=0.7$. } \label{F2}
\end{figure}

\begin{figure}[tbp]
\centering
\includegraphics[height=6.5cm,width=6.5cm]{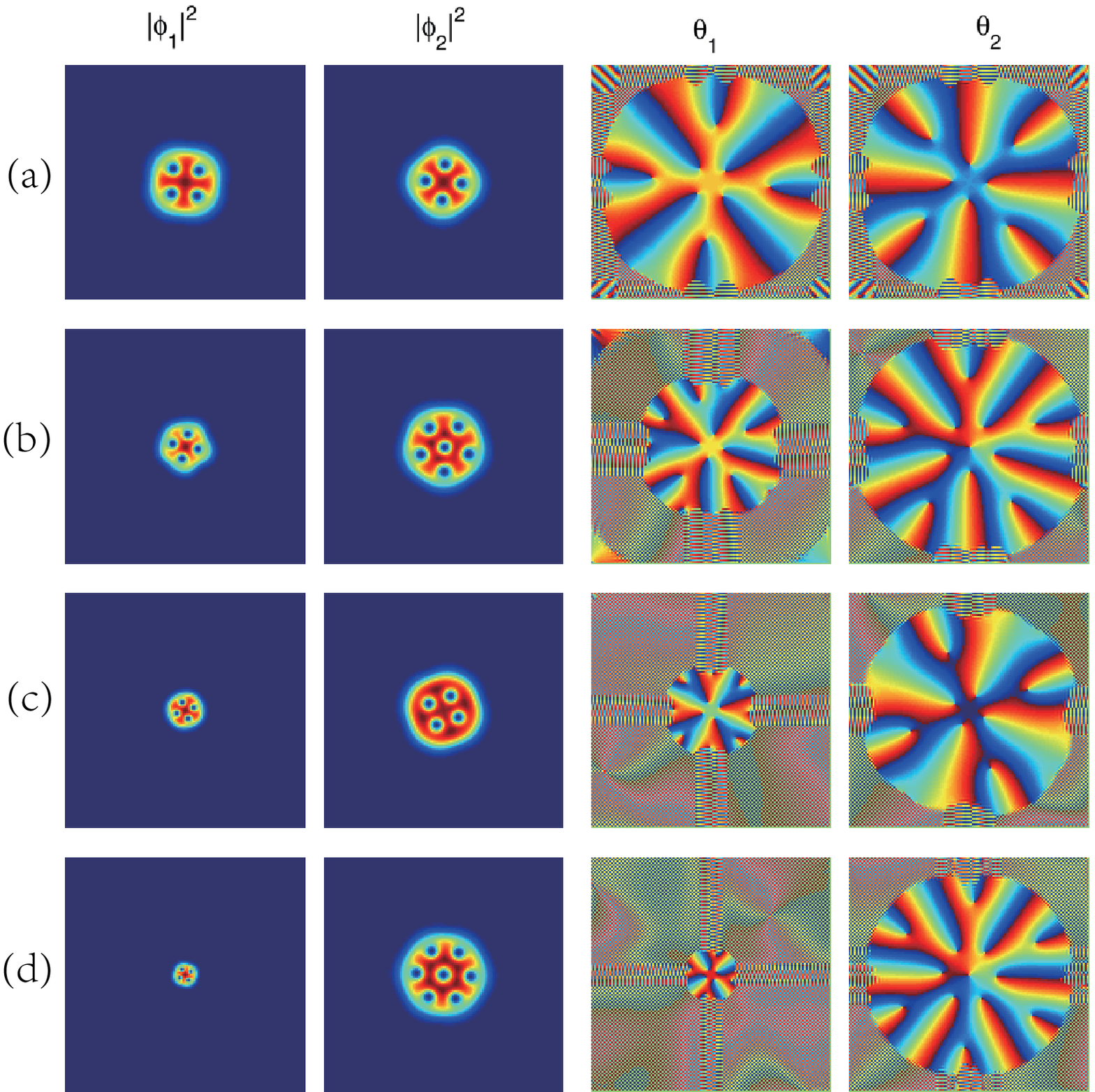}
\caption{(Color online) The same as in Fig. {\ref{F1}} but for $\Omega=0.8$. } \label{F3}
\end{figure}

\begin{figure}[tbp]
\centering
\includegraphics[height=6.5cm,width=6.5cm]{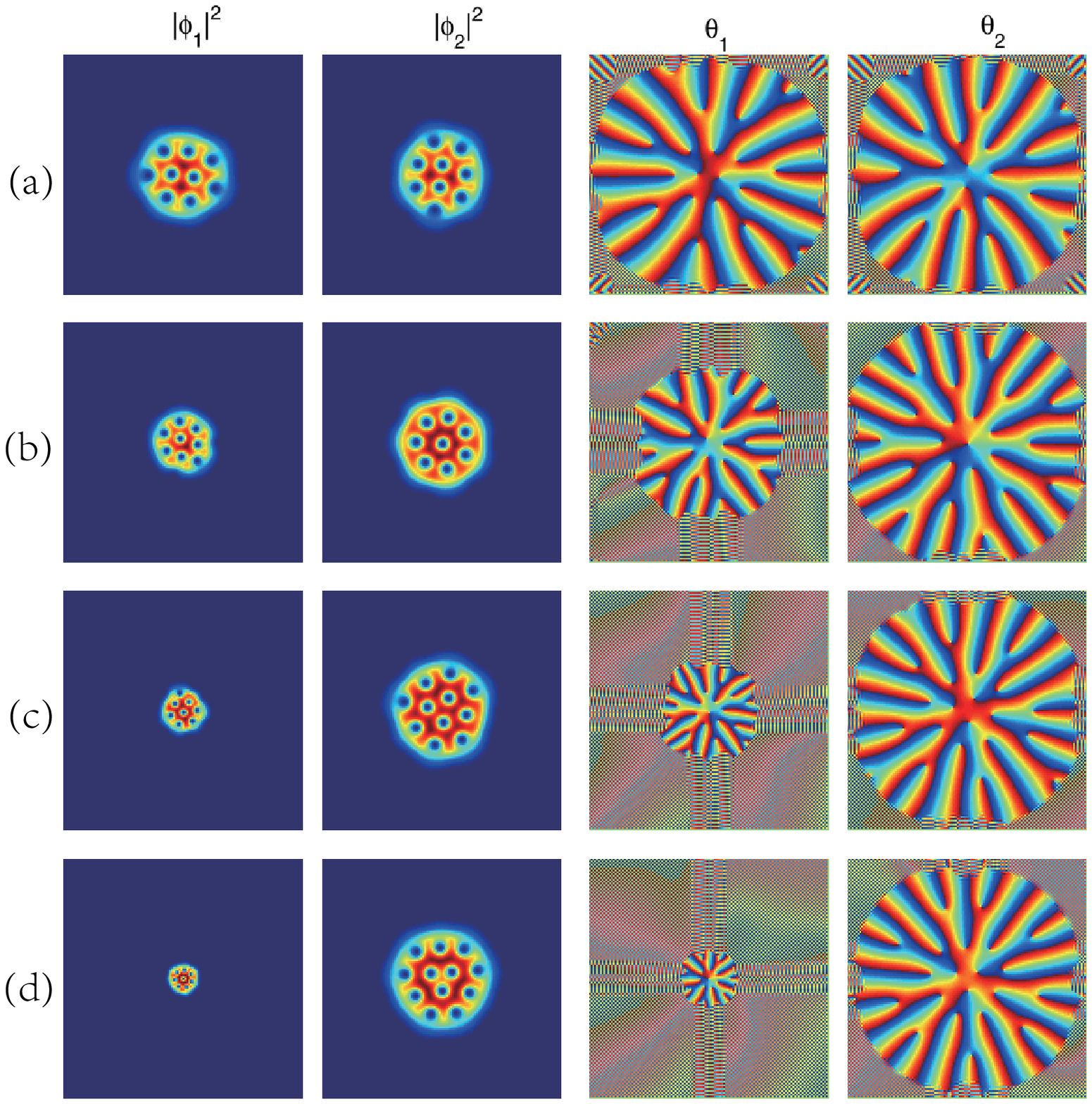}
\caption{(Color online) The same as in Fig. {\ref{F1}} but for $\Omega=0.9$. } \label{F4}
\end{figure}

Fig. 6. (Color online) The radius ratio of a binary rotating atomic BEC as a function of
the ratio of atomic masses. The contact interactions are fixed at $\textmd{g}=100$,
$\textmd{g}_{12}=20$, and the rotation frequency $\Omega=0.6,0.7,0.8,0.9$, corresponding to
(a), (b), (c), and (d) respectively. Here the mass of component $2$ is set as the unit mass, and the curve is
plotted based on the formula $R_1/R_2=[(11m^2_2+m_1m_2)/(11m^2_1+m_1m_2)]^{1/5}$. The blue rings correspond to the numerical results read
from Figs. {\ref{F1}-\ref{F4}} for $\gamma=1,2,4,9$, respectively.

\begin{figure}[tbp]
\centering
\includegraphics[height=5.0cm,width=9.5cm]{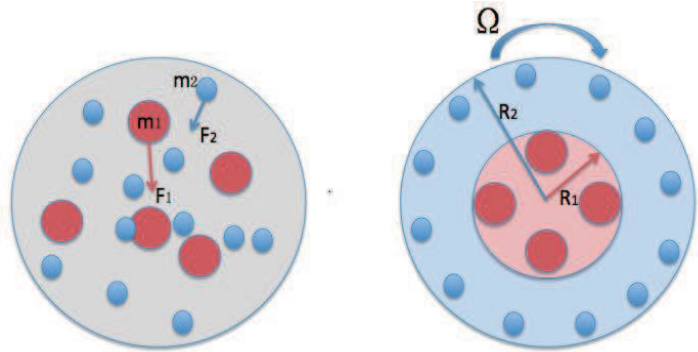}
\caption{(Color online) The qualitative explanation of the relation between the radius and the atomic mass of the condensate.
When one rotate a system mixing two species with unequal atomic masses with fixed
torque $\bf{\tau}=\textbf{r}\times \textbf{F}$,
where $\textbf{F}=m\bf{\Omega}\times(\bf{\Omega}\times \textbf{r})$ represents the centripetal force,
the bosons move as two clouds with different radius.} \label{F5}
\end{figure}

\begin{figure}[tbp]
\centering
\includegraphics[height=5.5cm,width=8.0cm]{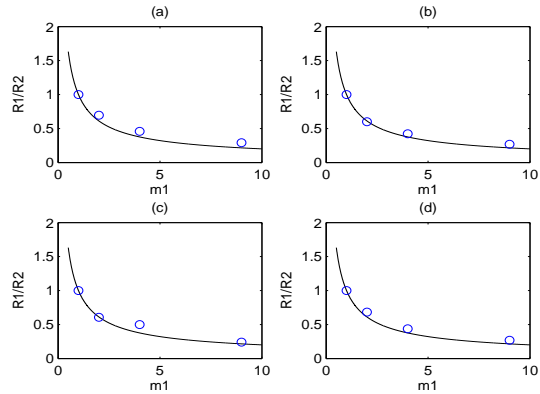}
\caption{(Color online) The radius ratio of a binary rotating atomic BEC as a function of
the ratio of atomic masses. The contact interactions are fixed at $\textmd{g}=100$,
$\textmd{g}_{12}=20$, and the rotation frequency $\Omega=0.6,0.7,0.8,0.9$, corresponding to
(a), (b), (c), and (d) respectively. Here the mass of component $2$ is set as the unit mass, and the curve is
plotted based on the formula $R_1/R_2=[(11m^2_2+m_1m_2)/(11m^2_1+m_1m_2)]^{1/5}$. The blue rings correspond to the numerical results read
from Figs. {\ref{F1}-\ref{F4}} for $\gamma=1,2,4,9$, respectively.} \label{F6}
\end{figure}

\begin{figure}[tbp]
\centering
\includegraphics[height=5.5cm,width=8.0cm]{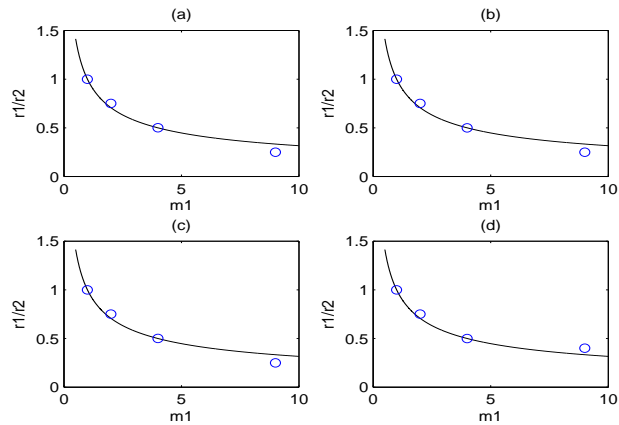}
\caption{(Color online) The size ratio of the core region for the vortices in the same condition
as in Fig. {\ref{F5}}. The curve is plotted based on the formula $r_1/r_2=m_1^{-1/2}$, and the blue
rings correspond to the results read from Figs. \ref{F1}-\ref{F4} for $\gamma=1,2,4,9$, respectively.} \label{F7}
\end{figure}

Fig. 7. (Color online) The size ratio of the core region for the vortices in the same condition
as in Fig. {\ref{F5}}. The curve is plotted based on the formula $r_1/r_2=m_1^{-1/2}$, and the blue
rings correspond to the results read from Figs. \ref{F1}-\ref{F4} for $\gamma=1,2,4,9$, respectively.

\newpage

\newpage

\newpage

\newpage

\newpage

\newpage

\newpage

\end{document}